\documentclass{IEEEcsmag}

\usepackage[colorlinks,urlcolor=blue,linkcolor=blue,citecolor=blue]{hyperref}

\usepackage{subcaption} 
\usepackage{upmath}
\usepackage{algorithm}
\usepackage{algpseudocode}
\usepackage{amsmath}
\usepackage{comment}

\jvol{XX}
\jnum{XX}
\paper{8}
\jmonth{May/June}
\jname{IT Professional}
\pubyear{2025}

\begin{document}

\sptitle{DEPARTMENT: Graphically Speaking}

\title{Hybridizing Expressive Rendering: Stroke-Based Rendering with Classic and Neural Methods}

\author{Kapil Dev}
\affil{RMIT University}


\markboth{GRAPHICALLY SPEAKING}{DEPARTMENT HEAD}

\begin{abstract}
\looseness-1 Non-Photorealistic Rendering (NPR) has long been used to create artistic visualizations that prioritize style over realism, enabling the depiction of a wide range of aesthetic effects, from hand-drawn sketches to painterly renderings. While classical NPR methods, such as edge detection, toon shading, and geometric abstraction, have been well-established in both research and practice, with a particular focus on stroke-based rendering, the recent rise of deep learning represents a paradigm shift. We analyze the similarities and differences between classical and neural network based NPR techniques, focusing on stroke-based rendering (SBR), highlighting their strengths and limitations. We discuss trade-offs in quality and artistic control between these paradigms, propose a framework where these approaches can be combined for new possibilities in expressive rendering.
\end{abstract}

\maketitle

\begin{figure}[t!]
    \centering
    \includegraphics[width=0.23\textwidth]{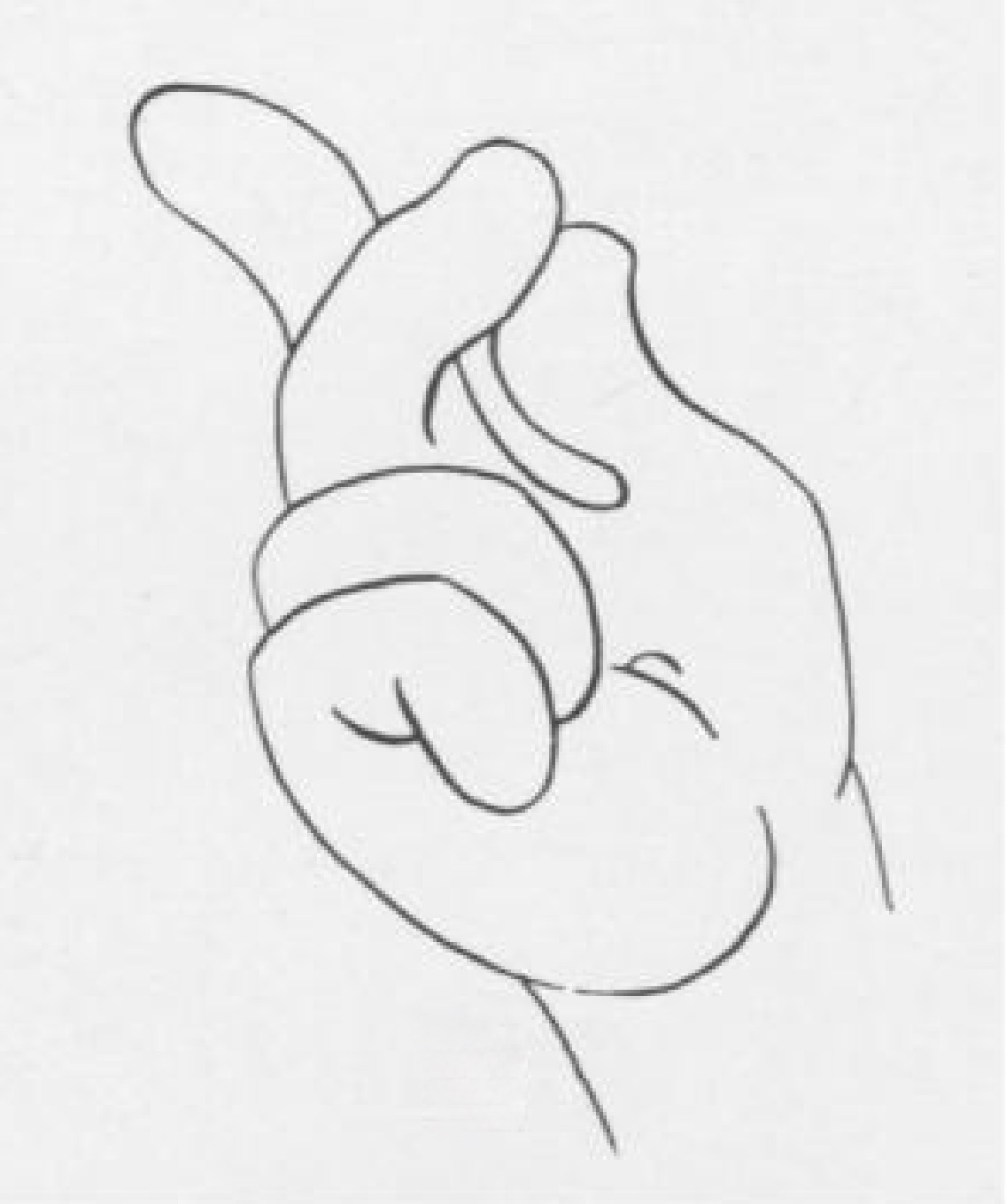}
    \includegraphics[width=0.23\textwidth]{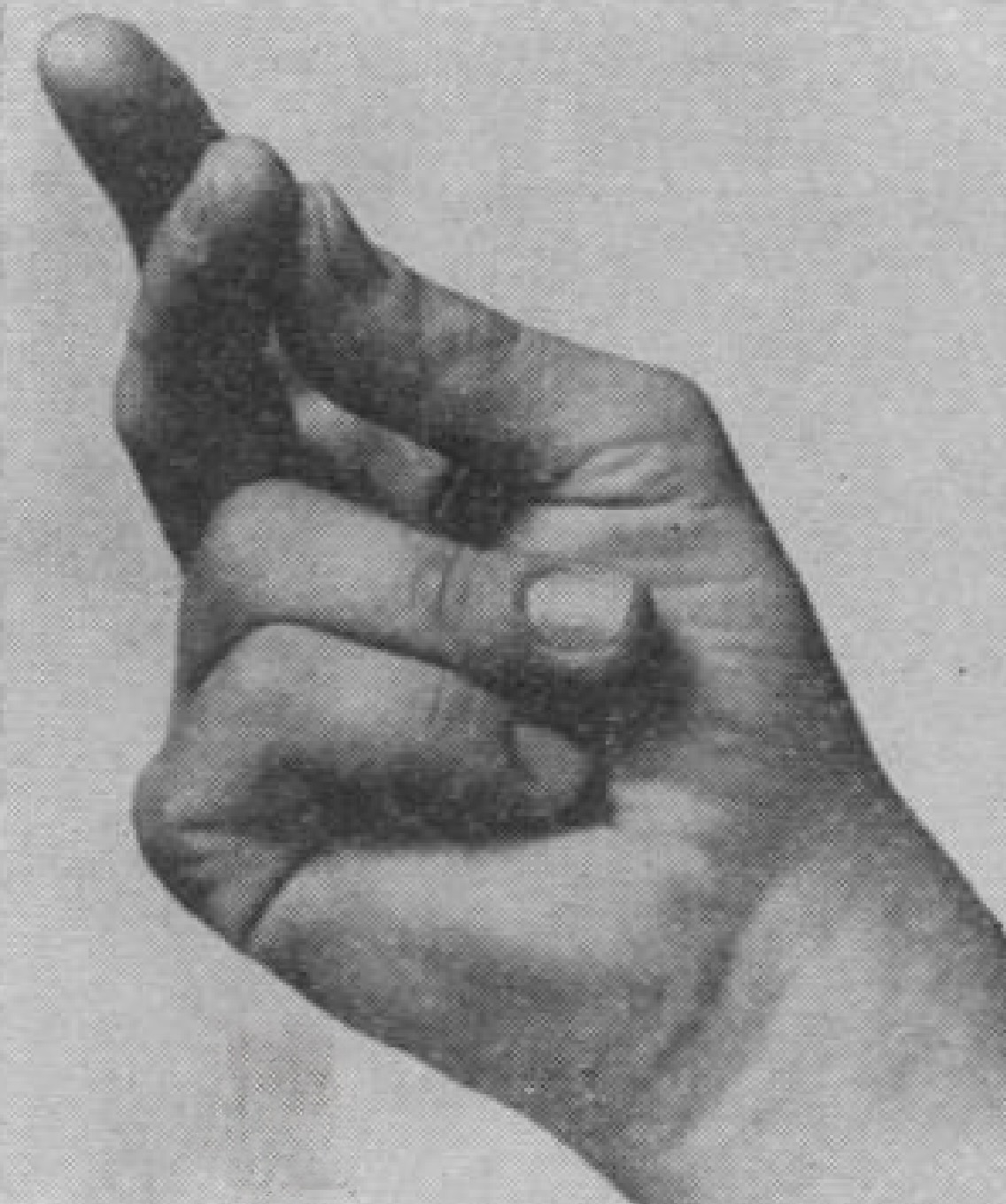}
    \caption{People respond faster and more accurately to drawings with exaggerated features (left) than they do to photographs (right). \textcopyright~1956 Ryan and Schwartz \cite{ryan1956speed}.}
    \label{fig:hand} 
\end{figure}

\chapteri{E}arly Non-Photorealistic Rendering (NPR) research focused on emulating traditional artistic mediums, such as sketching, watercolor, and oil painting, providing foundational insights into computational representation of artistic styles. Techniques like Haeberli’s image-based painting demonstrated how brush strokes could replicate Impressionist effects, illustrating the potential of algorithms to analyze and mimic artistic processes \cite{kyprianidis2012state}. As the field evolved, advanced methods like stippling, hatching, and line drawings emerged, emphasizing NPR’s utility in technical illustrations for improving clarity and communication \cite{kyprianidis2012state, hertzmann2003survey}. Indeed, as research by Ryan and Schwartz indicates, deliberately moving away from photographic realism, a core principle of NPR, can improve the speed and accuracy of visual perception \cite{ryan1956speed} (see Figure \ref{fig:hand}).

We use the term ``Classical NPR'' to describe techniques that rely on explicitly programmed rules and algorithms to mimic traditional artistic styles such as sketching, hatching, and painting. By contrast, we refer to ``Neural NPR'' as the application of deep learning models to achieve expressive renderings. Examples include neural style transfer for applying styles from reference images and generative adversatial networks for synthesizing images in specific artistic styles \cite{jing2019neural}.

The evolution of NPR marks a significant shift from handcrafted algorithms grounded in mathematical principles to data-driven models powered by deep learning. Classical or traditional techniques \cite{lansdown1995expressive}, such as silhouette rendering and tone shading, relied on explicitly defined geometric and visual rules. 
For example, stroke-based rendering, exemplified by Hertzmann’s seminal work, depends on geometric constraints to align strokes with image gradients, producing painterly effects but limited in stylistic variation \cite{hertzmann2003survey}.

\begin{figure*}[ht]
    \centering
    \includegraphics[width=0.42\textwidth]{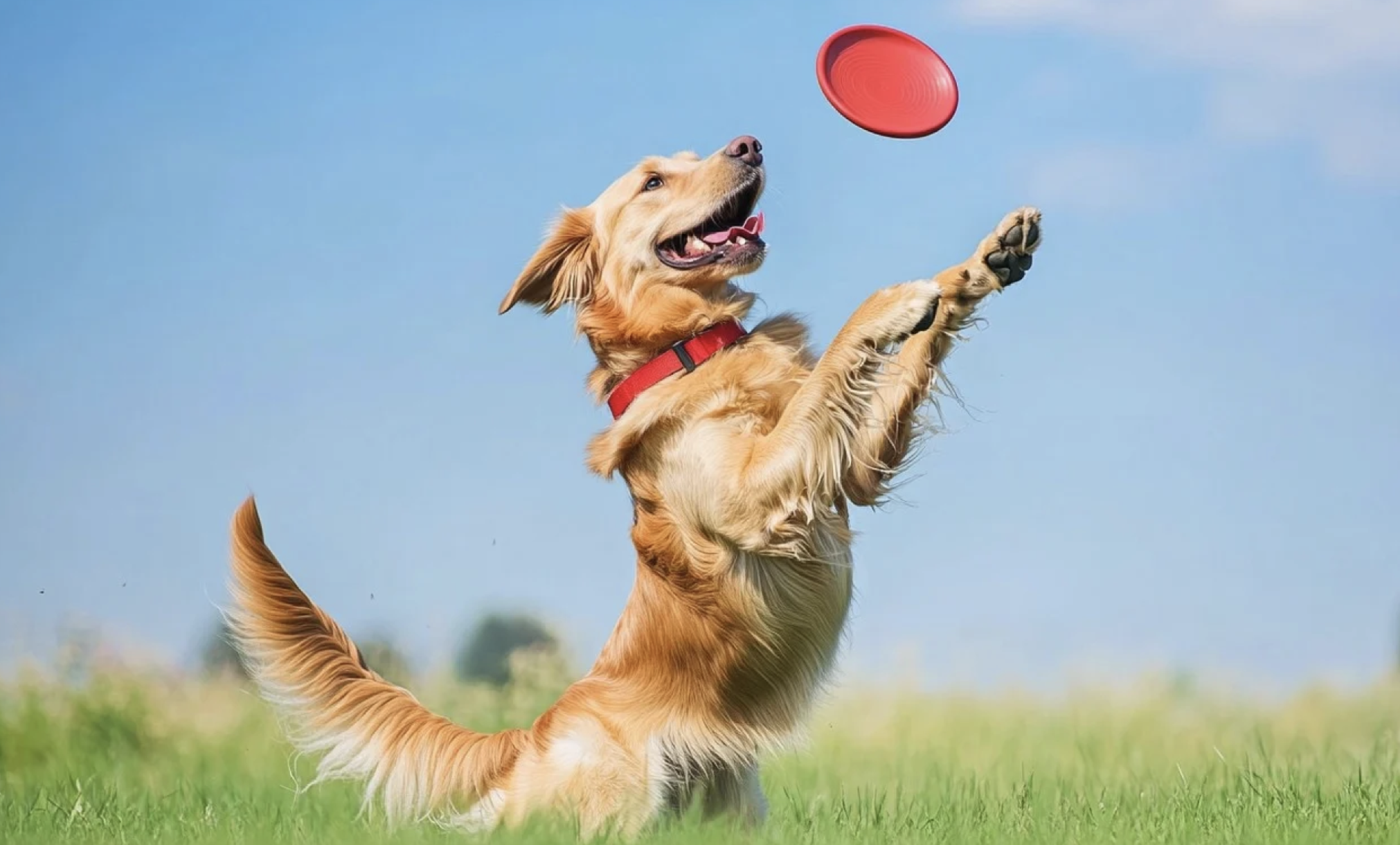}
    \includegraphics[width=0.56\textwidth]{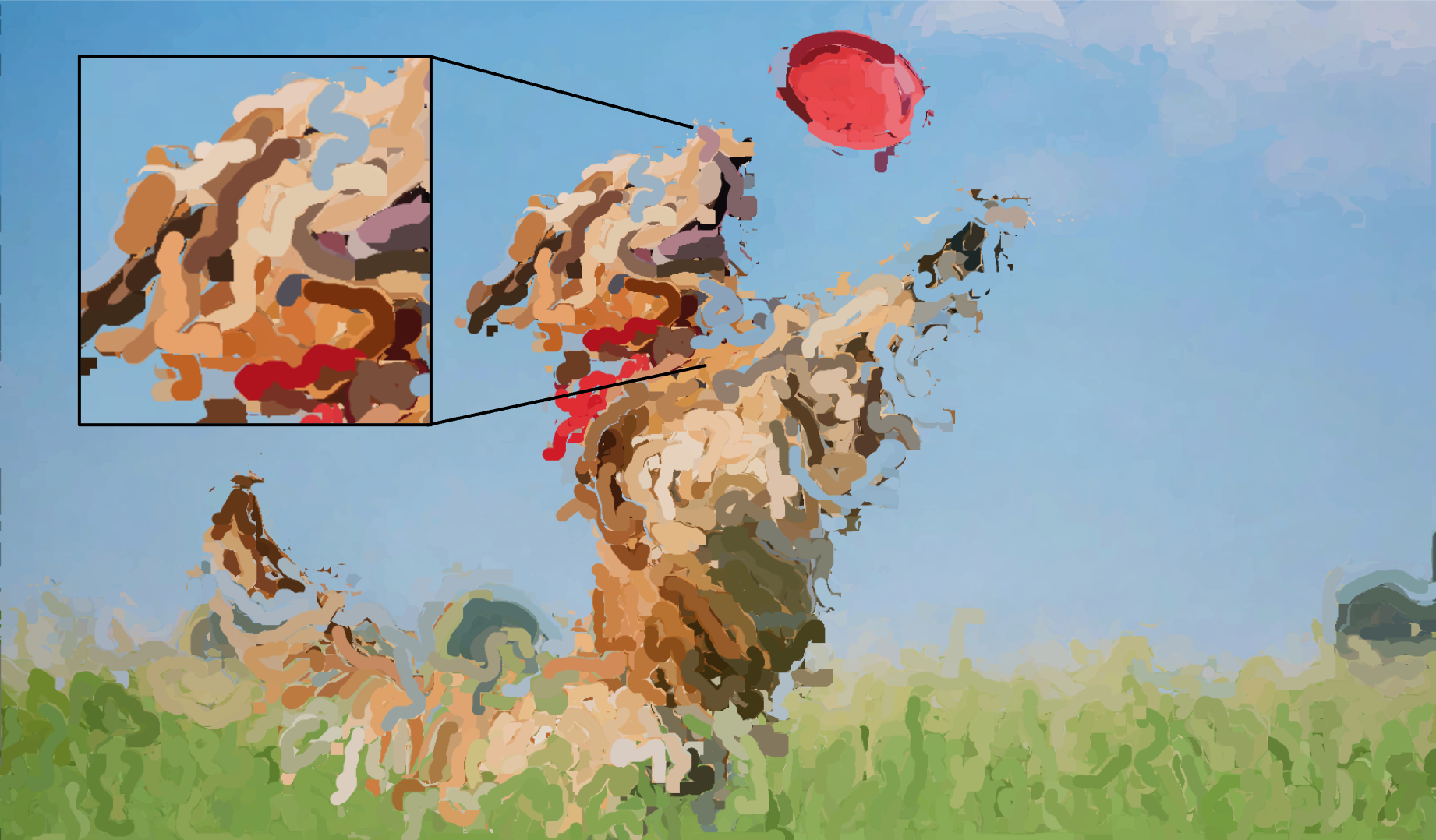}
    \caption{Left: Original image. Right: Intermediate rendering layer (Layer 1) with a specific brush stroke scale we implemented based on Hertzmann \cite{hertzmann2003survey}. Additional layers refine stroke detail to produce the final rendering (Algorithm \ref{alg:traditional_pipeline}). See the original paper for other layer examples.}
    \label{fig:fig2}
\end{figure*}

However, these deterministic algorithms, while interpretable, often lack the adaptability and stylistic diversity found in deep learning methods. Neural models, trained on extensive datasets, offer a paradigm shift. They learn implicit mappings from input data to stylized outputs, offering greater expressive potential \cite{jing2019neural}. Convolutional Neural Networks (CNNs) are commonly used to encode abstractions of texture and contours, facilitating generalization across styles.

\begin{algorithm}
\caption{Classical NPR Pipeline Example}
\label{alg:traditional_pipeline}
\begin{algorithmic}[t]
    \State \textbf{Input:} Natural Image $I$
    \State \textbf{Output:} Stylized Image $I_s$
    
    \State \textbf{// Feature Extraction}
    \State Compute gradient map: $\nabla I = \left[ \frac{\partial I}{\partial x}, \frac{\partial I}{\partial y} \right]$
    \State Extract edges: $E = \text{Threshold}(\nabla I)$

    \State \textbf{// Hand-Crafted Stylization}
    \State Apply toon shading: $I_t = \text{Quantize}(I, n)$ 
    
    \State Apply stroke-based rendering: $I_s = \sum_{i=1}^{N} S_i(E, I_t)$, where $S_i$ represents stroke placement and uses $I_t$ for color guidance.

    \State \textbf{Return:} Stylized image $I_s$
\end{algorithmic}
\end{algorithm}

Expanding NPR's expressive potential, John Lansdown and Simon Schofield authored the first comprehensive review of expressive renderings in 1995 \cite{lansdown1995expressive}. Our subsequent review focused specifically on expressive rendering on mobile devices \cite{dev2013mobile}, highlighting the challenges and opportunities presented by resource-constrained platforms. Aaron Hertzaman and a few others played a key role in development of this field \cite{hertzmann2003survey, hertzmann2010non}.

This paper discusses the key algorithmic ideas in both classical and neural NPR algorithms, and the level of artistic control and flexibility available to the users of these techniques. We also propose a hybrid framework where these approaches can be combined to explore possibilities to overcome their current limitations in stylistic control and speed.

\section{NPR Algorithm Development}
Early NPR research focused on techniques like toon shading and edge detection, relying on hand-crafted features that required significant domain expertise. However, the shift to data driven approaches, such as neural style transfer, has enabled more diverse artistic outputs by learning high-level artistic representations directly from reference images.

\subsection{Stroke-Based Methods}
Advancements in stroke-based rendering (SBR) have significantly influenced NPR \cite{hertzmann2003survey, lansdown1995expressive}. These methods use algorithms to place strokes, much like a painter deciding where to add each brushstroke (Figure \ref{fig:fig2}). This decision-making process can be represented with a function: $Stroke_{n+1} = f(Image_n, Goal)$, where $Stroke_{n+1}$ is the next stroke to be added, $Image_n$ is the current state of the image, and $Goal$ represents the desired final image (Algorithm \ref{alg:traditional_pipeline}). The function $f$ represents the algorithm or agent that determines the best stroke to add based on the current image and the goal. Early SBR methods, such as Haeberli's paint program, were semi-automatic and required user input for stroke placement. Subsequent research introduced automation techniques like image gradient analysis and curve tracing. Modern SBR algorithms leverage deep reinforcement learning to train agents capable of rendering images autonomously, stroke by stroke \cite{liu2021paint}.

\subsection{Vision-Based Methods}
The integration of computer vision techniques marked another turning point in NPR development. Vision-Based Methods use computer algorithms to `see' and `analyze' images, extracting features that guide artistic rendering. We can represent this process with a simple function: $Features = g(Image)$, where $Image$ is the input image, and $Features$ are the extracted characteristics, such as edges, textures, or salient regions. The function $g$ represents the vision algorithm that performs this extraction, allowing the system to understand and interpret the image's content for artistic purposes. Early methods relied on low-level image processing, such as the Sobel operator, to detect edges and features \cite{lansdown1995expressive}. Over time, higher-level techniques, including image segmentation and salience detection, were incorporated to guide artistic rendering, enabling more perceptually-driven and context-aware styles. Beyond extracting key visual features, style transformation techniques emerged to modify these extracted elements, leading to more artistic flexibility.
\begin{figure*}[t]
    \centering
    \includegraphics[width=0.49\textwidth]{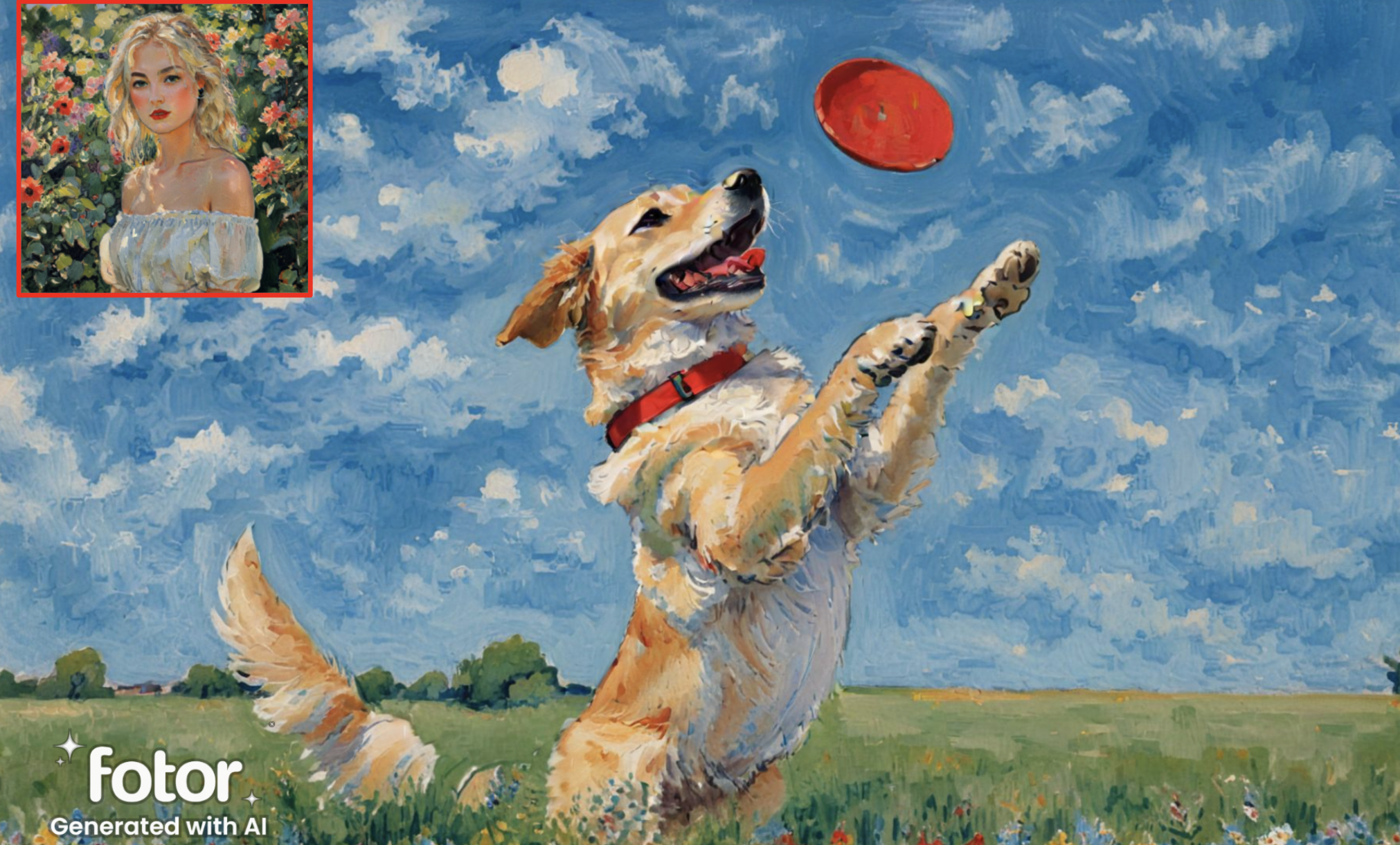}
    \includegraphics[width=0.49\textwidth]{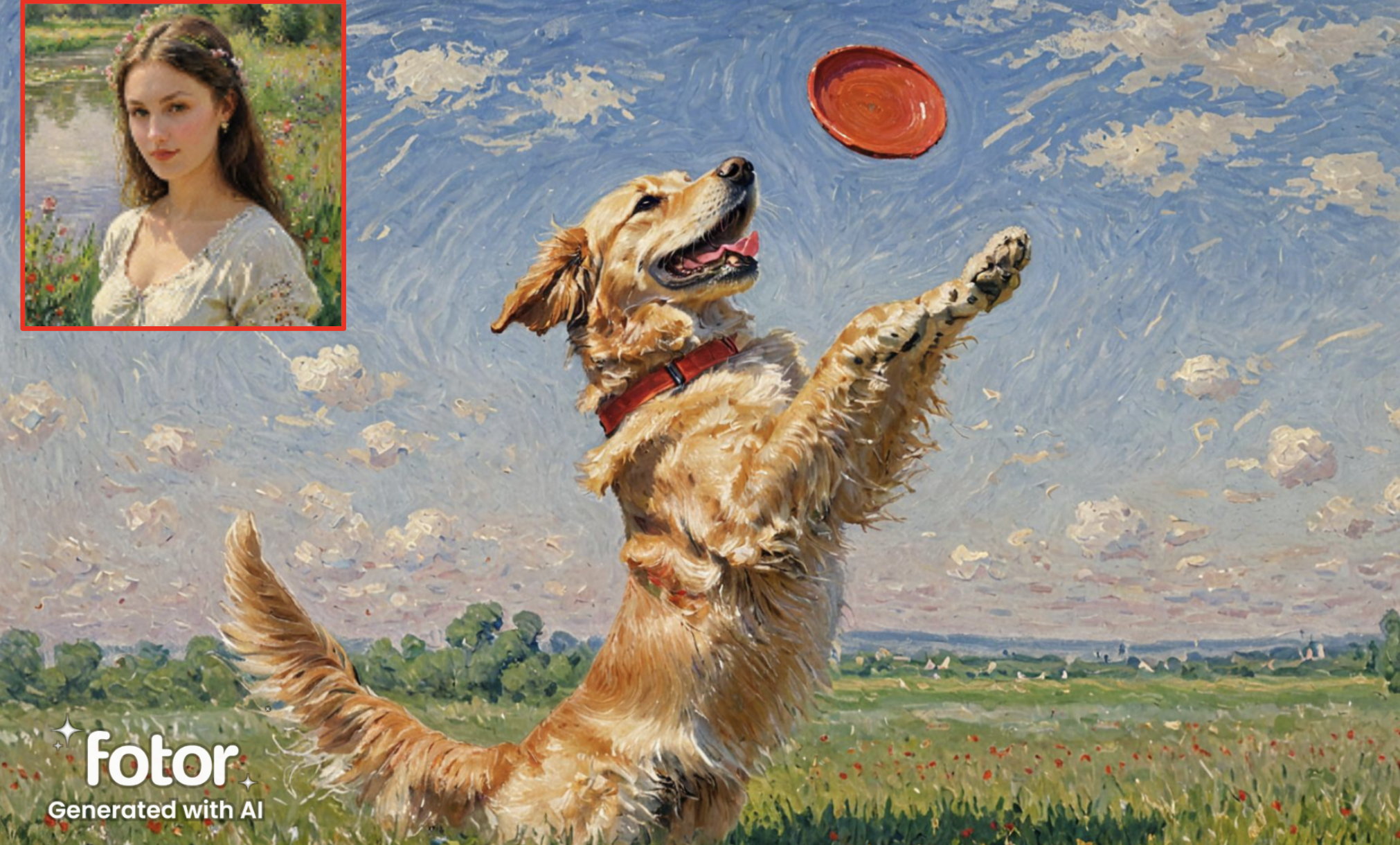}
    \caption{This figure shows two stylized renderings of the original content image in Figure \ref{fig:fig2}, created using Fotor's GoArt (https://goart.fotor.com/). The style images are located in the top left corners of the stylized results. See algorithm \ref{alg:neural_pipeline} for a general pipeline for creating such images.}
    \label{fig:dog_neural}    
\end{figure*}

\begin{algorithm}
\caption{Neural Pipeline}
\label{alg:neural_pipeline}
\begin{algorithmic}[1]
    \State \textbf{Input:} Content Image $I_c$, Style Image $I_s$
    \State \textbf{Output:} Stylized Image $I_t$
    
    \State \textbf{// Feature Extraction Using CNN (i.e. $\phi$)}
    \State Extract content features: $F_c = \phi(I_c)$
    \State Extract style features: $F_s = \phi(I_s)$
    
    \State \textbf{// Style Transfer via Optimization}
    \State Initialize $I_t = I_c$ (or random noise)
    
    \For{each iteration}
        \State Compute content loss:
        \[
        \mathcal{L}_{content} = ||F_c - \phi(I_t)||^2
        \]
        \State Compute style loss using Gram matrices:
        \[
        G_s = F_s F_s^T, \quad G_t = \phi(I_t) \phi(I_t)^T
        \]
        \[
        \mathcal{L}_{style} = \sum_l ||G_s^l - G_t^l||^2
        \]
        \State Compute total loss:
        \[
        \mathcal{L} = \alpha \mathcal{L}_{content} + \beta \mathcal{L}_{style}
        \]
        \State Update $I_t$ via gradient descent:
        \[
        I_t \leftarrow I_t - \eta \frac{\partial \mathcal{L}}{\partial I_t}
        \]
    \EndFor
    
    \State \textbf{Return:} Stylized image $I_t$
\end{algorithmic}
\end{algorithm}

\subsection{Optimization in NPR}
Optimization algorithms have also played a pivotal role in NPR's progress. 
Optimization in NPR aims to find the `best' possible rendering by minimizing an error or maximizing a quality score. This can be represented as: $Best\ Rendering = argmin\ Error(Rendering)$ or $Best\ Rendering = argmax\ Quality(Rendering)$, where $Rendering$ represents a possible output, $Error(Rendering)$ is a measure of how far it is from the desired result, and $Quality(Rendering)$ is a measure of how good it is. $argmin$ finds the $Rendering$ that minimizes the error, and $argmax$ finds the $Rendering$ that maximizes the quality.

Classical methods optimize artistic goals within strict constraints, such as minimizing the number of strokes required to represent an image while maintaining perceptual fidelity, or aligning strokes with salient edges to enhance visual impact. These methods often rely on pixel-wise metrics like L2 norms to quantify differences between inputs and rendered outputs while enforcing sparsity constraints to achieve painterly effects. Although precise, these approaches face scalability challenges as the complexity of artistic goals increases.

In contrast, deep learning optimizes objectives such as style loss and content preservation through backpropagation in neural networks, bypassing the need for explicit objective formulations. Style loss, for example, might involve matching the Gram matrices of feature activations between a stylized output and a reference style image, capturing the texture and color characteristics of the desired style. 

\subsubsection{Neural Image Stylization} These methods leverage techniques to alter an image's visual style using either a reference-style image or a dataset of styles (Figure \ref{fig:dog_neural}). They operate holistically on the image, without explicitly simulating the step-by-step artistic process. Neural Style Transfer works by combining the `content' of one image with the `style' of another. We can represent this idea with a simple equation: $Generated\ Image = \alpha \times Content + \beta \times Style$, where $Content$ represents the visual structure of one image, $Style$ represents the artistic characteristics of another, and $\alpha$ and $\beta$ are weights that control how much of each is included in the $Generated\ Image$. By adjusting $\alpha$ and $\beta$, we can fine-tune the resulting style transfer (Algorithm \ref{alg:neural_pipeline}).

Generative Adversarial Networks (GANs), including CycleGAN, AttentionGAN, and Gated-GAN, extend this capability by training on large datasets of artistic styles, enabling transformations that do not rely on a specific reference image. Although these methods excel in generating visually appealing results, they often lack control over fine-grained details, making them less suitable for tasks that require procedural or interactive customization.

\subsubsection{Stoke Level Neural Techniques}
These techniques build images step-by-step, like a painter adding brushstrokes \cite{jing2019neural}. 
This category draws inspiration from traditional art techniques, with algorithms like Sketch-RNN employing recurrent neural networks (RNNs) to convert input images into sequential line drawings. Methods for replicating oil and watercolor paintings use reinforcement learning or differentiable rendering techniques to guide the placement of strokes, creating images with rich textures and depth. However, these methods face challenges in achieving the spontaneity and unpredictability that characterize human creativity.

\begin{table*}[ht!]
\centering
\caption{Comparison of Artistic Control in Classical and Neural NPR}
\begin{tabular}{|c|c|c|}
\hline
\textbf{Feature} & \textbf{Classical NPR (SBR, Toon Shading)} & \textbf{Neural NPR (NST, Transformers, GANs)} \\
\hline
\textbf{Stroke Placement} & Explicit, rule-based (gradients etc.) & Implicit, learned via deep models (attention, priors) \\
\hline
\textbf{Customization} & High, user-defined stroke parameters & Moderate, via hyperparameters or embeddings \\
\hline
\textbf{Adaptability} & Limited, manual tuning per style & Learns styles dynamically from data \\
\hline
\textbf{Computational Cost} & Low to moderate, efficient in real-time & High; optimized via feed-forward models \\
\hline
\textbf{Size} (\(s\)) & User-defined or procedural & Learned via filters, feature scaling \\
\hline
\textbf{Orientation} (\(\theta\)) & Edge-aligned, user-guided & Adjusted via feature maps, attention \\
\hline
\textbf{Thickness} (\(t\)) & Controlled by pressure, heuristics & Learned via loss functions, filters \\
\hline
\textbf{Color} (\(c\)) & Manual or rule-based assignment & Extracted via Deep Learning \\
\hline
\textbf{Texture} (\(\tau\)) & Procedural (stippling, hatching) & Neural texture synthesis (Gram, CNNs) \\
\hline
\textbf{Opacity} (\(o\)) & User-defined transparency & Adaptive blending via learned masks \\
\hline
\textbf{Stroke Coherence} & High, direct placement control & Variable, improves with attention models \\
\hline
\textbf{Temporal Coherence} & Optical flow, manual tracking & GANs, loss-based stabilization \\
\hline
\textbf{Suitability} & Precise control for manual design & Scalable, automated stylization \\
\hline
\end{tabular}
\label{tab:NPR_comparison}
\end{table*}

\section{Stroke-Level Control in NPR}
The foundational role of SBR in classical NPR highlights manual stroke placement optimization. This serves as a baseline for comparing neural models that aim to emulate similar styles with learned representations \cite{hertzmann2003survey}. Unlike pixel-based methods, stroke-based approaches offer greater control and intuitiveness for artistic image creation. The transition from classical techniques, such as image analogies, region-based abstraction, and image filtering, to deep learning-based methods has transformed the field \cite{jing2019neural}.

Classical Non-Photorealistic Rendering (NPR) techniques primarily focus on image reconstruction into non-photorealistic styles through Stroke-Based Rendering (SBR). These methods rely on algorithms that decompose images into discrete stroke elements and govern their rendering.
SBR algorithms commonly employ a greedy approach to decompose strokes into constituent steps, necessitating considerable manual intervention. Hertzmann introduced a style design method utilizing spline brushstrokes, where the manipulation of parameters dictates the resultant painting effects, shows the necessity for user expertise in guiding the drawing process \cite{hertzmann2003survey}.
They depend heavily on algorithms that decompose images into strokes and dictate how those strokes are rendered. These methods typically produce a restricted range of styles, often based on the specific algorithm employed. Many SBR methods require substantial user input to guide stroke placement, shape selection, and parameter adjustments. These methods can be slow and computationally intensive, particularly when dealing with complex images.

Deep learning models, particularly in image generation and style transfer incorporate perceptual measures by using perceptual loss functions instead of relying solely on pixel-wise errors. These functions are computed using pre-trained deep neural networks, which have learned to extract perceptually meaningful features from large image datasets. 
A common approach uses features from a pre-trained VGG network, where the activations from intermediate layers capture different levels of content and style abstraction. By comparing these features between a generated and target image, a perceptual loss function can be defined. In style transfer, this involves extracting VGG features, computing Gram matrices to capture style correlations, and calculating style loss by comparing the Gram matrices of the generated and reference images, encouraging similar perceptual characteristics \cite{jing2019neural}. Unlike typical NST that uses texture patches, authors in \cite{liu2023painterly} learn stroke patterns from real paintings, allowing outputs with a more authentic, ``painterly'' feel by applying style through realistic brushstrokes.

\begin{figure}[t!]
    \centering
    \includegraphics[width=0.48\textwidth]{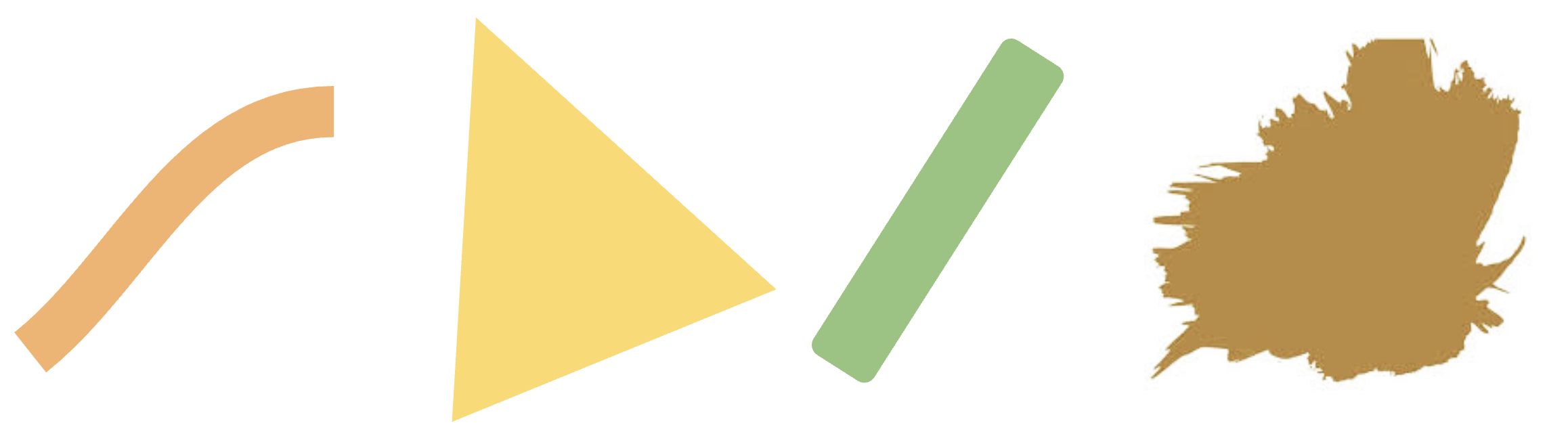}
    \caption{Example brush stroke models used in stroke based non-photorealistic rendering (left to right: curved, triangle, rectangle and random raster, all drawn by hand for demonstration). For example, the curved brush stroke model is used in the Figure 2 above.}
    \label{fig:brushes} 
\end{figure}

\subsection{Artistic Control in Stroke-Based Rendering}
SBR is fundamental to artistic rendering, where strokes act as primary rendering primitives. Each stroke is characterized by:  
\begin{itemize}
    \item \textbf{Size} (\(s\)): Defines stroke coverage, affecting detail preservation.
    \item \textbf{Orientation} (\(\theta\)): Determines alignment with image features.
    \item \textbf{Thickness} (\(t\)): Controls stroke weight and prominence.
    \item \textbf{Color} (\(c\)): Sets hue, saturation, and brightness.
    \item \textbf{Texture} (\(\tau\)): Influences roughness, smoothness, or blending.
    \item \textbf{Opacity} (\(o\)): Adjusts transparency for layering and depth.
\end{itemize}

Classical NPR methods \cite{hertzmann2003survey}, define stroke attributes using hand-crafted rules and user input, allowing fine-grained artistic control but requiring manual tuning. In contrast, data-driven methods, including NST and GAN-based models, automate stroke generation by learning stylistic features from data (Table~\ref{tab:NPR_comparison}). 

Paint Transformer introduces a paradigm shift in data-driven stroke-based rendering by employing a feed-forward transformer for parallel stroke prediction, significantly improving speed and efficiency over traditional methods \cite{liu2021paint}. Specifically, it predicts stroke parameters—position, curvature, and opacity—via a transformer network, enabling scalable rendering but reducing direct user intervention. While neural NPR excels in efficiency and adaptability, it often sacrifices precise user-driven stroke control.

NST applies an artistic effect to an image by balancing content preservation and style adaptation. This is achieved by minimizing a combined loss function:
\begin{equation}
    \mathcal{L} = \alpha \mathcal{L}_{content} + \beta \mathcal{L}_{style},
\end{equation}
where \(\alpha\) and \(\beta\) determine the trade-off between maintaining image structure and applying the desired style. Content loss ensures that the stylized image retains recognizable features of the original:
\begin{equation}
    \mathcal{L}_{content} = ||F_c - F(I_t)||^2,
\end{equation}
where \(F_c\) represents extracted content features, and \(F(I_t)\) represents features of the generated image.

Style loss transfers artistic patterns by comparing texture statistics using Gram matrices:
\begin{equation}
    G_s = F_s F_s^T, \quad G_t = F(I_t) F(I_t)^T.
\end{equation}
The closer \(G_t\) is to \(G_s\), the better the stylization matches the reference style.

Advanced NST techniques enhance stroke-level control by integrating adaptive normalization, 
content-aware transformations, and dynamic parameterization strategies. 
For example content-aware transformations further improve structural alignment by 
geometric constraints to guide stroke orientation. While standard NST does not explicitly modify the content feature 
map $F(I)$ to control stroke direction, recent extensions introduce edge-aware mechanisms and spatial attention to 
ensure strokes align with dominant image structures \cite{jing2019neural, 
hertzmann2003survey}. 

While these improvements provide some stroke-level control, NST still lacks explicit per-stroke manipulation, making it less interactive compared to traditional NPR methods.

\begin{algorithm}[t]
\caption{Hybrid Stroke Planning}
\label{alg:hybrid_stroke_general}
\begin{algorithmic}[1]
\Require Input image or scene $I$, Deep model $\mathcal{M}$
\Ensure Optimized stroke sequence $\mathcal{S}^*$ for rendering

\State \textbf{Step 1: Preprocessing and Feature Extraction}
\State Extract edges, contours, and saliency etc. from $I$
\State Generate initial strokes $\mathcal{S} = \{s_1, s_2, \dots, s_n\}$

\State \textbf{Step 2: Rule-Based Initialization}
\For{each stroke $s_i \in \mathcal{S}$}
    \State Compute attributes (e.g., position, size)
    \State Apply heuristic constraints (e.g., edge alignment, density control)
    \State Assign priority based on saliency and region importance
\EndFor

\State \textbf{Step 3: Neural Refinement}
\For{each stroke $s_i \in \mathcal{S}$}
    \State Predict refined stroke parameters using $\mathcal{M}$
    \If{AI prediction conflicts with heuristic constraints}
        \State Apply correction to ensure coherence
    \EndIf
\EndFor

\State \textbf{Step 4: Hybridization}
\For{each stroke $s_i \in \mathcal{S}$}
    \State Blend neural and classical stroke decisions
\EndFor

\State \textbf{Step 5: Rendering and Post-Processing}
\State Render final stroke sequence $\mathcal{S}^*$
\State Apply post-processing (e.g., for coherence)

\State \Return Final stylized rendering
\end{algorithmic}
\end{algorithm}

\section{Framework for Hybrid Approaches}
Stroke sequence planning is a fundamental challenge in NPR, directly influencing the perceptual quality, structural coherence, and computational efficiency of stylized imagery. We propose a generalized hybrid framework for combining explicit heuristic control with adaptive neural refinement (see Algorithm \ref{alg:hybrid_stroke_general}.) 
The integration of rule-based heuristics and data-driven optimization in stroke sequence planning presents a scalable and adaptable framework for NPR.

\subsection{Step 1: Preprocessing}
This is the first step in the algorithm \ref{alg:hybrid_stroke_general} above. Stroke-based NPR requires guidance to ensure meaningful stroke placement. Unlike raw pixel manipulation, stroke-based methods benefit from explicit feature extraction, which provides spatial and structural information for stroke rendering. Pre-processing could help extract key visual characteristics such as \textit{ edges, saliency, and regional segmentation}, ensuring that strokes align with significant image structures. We present some approaches below on how to perform this step.

\textbf{Edge and Contour Detection:} Edges and contours serve as primary guides for stroke placement, particularly in painterly and sketch-based rendering. Edge detection techniques such as Sobel and Canny filters identify high-gradient regions, ensuring that strokes follow object boundaries. For example, in portrait rendering, edge detection preserves facial contours, leading to structured and visually coherent stroke layouts.

\textbf{Finding Salient Maps:} Saliency maps help prioritize regions of visual importance, ensuring that strokes emphasize perceptually significant areas. Deep learning-based saliency models, such as U-Net-based detectors, compute saliency maps $S(I)$ by predicting fixation points within an image. This is particularly useful in artistic rendering, where highly salient areas (e.g., eyes, lips in portraits) require finer stroke detailing, while less salient areas (e.g., background) may use coarser strokes.

\textbf{Stroke Density Estimation:} Uniform stroke distribution can lead to unnatural textures, necessitating adaptive stroke density estimation. Voronoi-based segmentation and clustering techniques ensure that stroke placement follows localized texture variations. This process partitions the image into non-overlapping regions, allowing for controlled stroke placement while preserving structural integrity.

\textbf{Generating Initial Stroke Candidates:} We can combine the above three or additional preprocessing steps to generate an initial stroke set. For example, edge, saliency, and density maps enables the generation of a structured stroke candidate set, $\mathcal{S} = \{s_1, s_2, \dots, s_n\}$, such that each candidate stroke is assigned an initial weight based on feature contributions:
\begin{equation}
W(s_i) = \alpha E(I) + \beta S(I) + \gamma D(I),
\end{equation}
where $E(I)$, $S(I)$, and $D(I)$ represent edge strength, saliency, and stroke density, respectively, and $\alpha, \beta, \gamma$ are weighting coefficients.

The preprocessing stage could output a structured set of candidate strokes, prioritized on the basis of visual importance. This provides a foundation for subsequent rule-based initialization and data-driven refinement with both structural and artistic elements.

\subsection{Step 2: Stroke Initialization}
This is the second step ``Step 2: Rule-Based Initialization'' in the algorithm \ref{alg:hybrid_stroke_general} we proposed above. Once stroke candidates $\mathcal{S}$ are generated, rule-based constraints refine their attributes for structured placement. While neural methods can introduce expressiveness, heuristics enforce essential artistic principles such as \textit{contour alignment, density control, and priority assignment}. This step ensures that strokes maintain structural integrity before data-driven refinement. We offer method suggestions below, and encourage readers to develop their own.

\textbf{Orientation Alignment:} Aligning strokes with local contours and image gradients ensures natural stroke placement. Given an edge map $E(I)$, stroke orientation $\theta_i$ is computed as:
\begin{equation}
\theta_i = \tan^{-1} \left( \frac{\partial I}{\partial y}, \frac{\partial I}{\partial x} \right).
\end{equation}
This alignment allows strokes to follow high-gradient areas, as seen in techniques like curvature-based hatching. For example, in landscape rendering, strokes align with terrain features to create depth and texture consistency.

\textbf{Stroke Density Control:} Unregulated stroke placement can lead to cluttered regions or overly sparse textures. To enforce density constraints, a Voronoi-based approach partitions the image into adaptive stroke regions, ensuring even coverage. The stroke density function $D(I)$ modulates the number of strokes per unit area:
\begin{equation}
D(I) = \frac{1}{Z} \sum_{i} W(s_i),
\end{equation}
where $W(s_i)$ represents the stroke weight derived from saliency and edge strength, and $Z$ is a normalization factor. This prevents oversaturation in high-detail areas while preserving texture in smooth regions.

\textbf{Saliency-Guided Stroke Prioritization:} Not all strokes contribute equally to perception. High-priority strokes are retained, while lower-priority ones may be removed or adjusted. Priority is assigned based on saliency $S(I)$:
\begin{equation}
P(s_i) = \lambda S(I) + (1 - \lambda) E(I),
\end{equation}
where $\lambda$ controls the relative influence of saliency versus edge strength. In portrait rendering, this ensures that facial features receive higher detail, while backgrounds use simplified strokes.

This step outputs a refined stroke set $\mathcal{S'}$ with optimized orientation, density, and priority. These constraints set a balance between structural accuracy and artistic flexibility, for neural refinement in the next stage.

\subsection{Step 3: Stroke Refinement}
While rule-based stroke initialization ensures structural consistency, it lacks adaptability to diverse artistic styles. Neural (or data driven) refinement addresses this limitation by learning stroke attributes from data. Using deep learning models such as transformers and reinforcement learning (RL), strokes are dynamically adjusted for improved style consistency and expressiveness. The following neural techniques can be used individually or combined, and readers are encouraged to consider other suitable approaches.

\textbf{Neural Stroke Prediction:} A learned model $\mathcal{M}$ predicts stroke refinements based on the initial stroke set $\mathcal{S'}$. Given an input stroke $s_i$ with attributes $(x_i, y_i, \theta_i, l_i, w_i)$, the AI model predicts an optimized stroke:
\begin{equation}
\hat{s_i} = \mathcal{M}(s_i) = (x'_i, y'_i, \theta'_i, l'_i, w'_i),
\end{equation}
where adjustments allow style adaptation while preserving coherence with image features. Transformer-based models are suited for encoding global stroke dependencies, improving consistency across complex scenes.


\textbf{Hybrid Correction:} While neural models introduce flexibility, they may generate artifacts or misaligned strokes. To mitigate this, we argue that a hybrid correction mechanism or similar approaches can be used to evaluate neural predictions against rule-based constraints:
\begin{equation}
\hat{s_i}^* = \text{blend}(\hat{s_i}, s_i) = \gamma \hat{s_i} + (1 - \gamma) s_i,
\end{equation}
where $\gamma$ determines the deep neural network's influence relative to heuristic rules. This step ensures deep model refined strokes remain structurally coherent.

Finally, this stage outputs a refined stroke sequence $\mathcal{S}^*$, balancing data-driven expressiveness with rule-based stability. These strokes are then finalized through hybrid decision fusion and rendering.

\subsection{Step 4: Hybridization}
While data-driven stroke refinement enhances adaptability, uncontrolled neural modifications may introduce artifacts or distort the underlying structure. This stage addresses this by balancing neural stroke modifications with rule-based heuristics, ensuring consistency and preserving perceptual coherence.

To maintain both structure and expressiveness, model-refined strokes $\hat{s_i}$ and rule-based strokes $s_i$ are adaptively blended:
\begin{equation}
s_i^* = \gamma \hat{s_i} + (1 - \gamma) s_i,
\end{equation}
where $\gamma \in [0,1]$ controls the deep neural network's influence. Higher $\gamma$ values prioritize learned artistic styles, while lower values enforce structural constraints.

\textbf{Perceptual Consistency Scoring:} To evaluate stroke coherence, we can use a scoring function. For example $Q(s_i)$ ranks strokes based on feature preservation and adherence to the original image:
\begin{equation}
Q(s_i) = \alpha S(s_i) + \beta E(s_i) - \lambda D(s_i),
\end{equation}
where $S(s_i)$ represents saliency alignment, $E(s_i)$ measures edge consistency, and $D(s_i)$ penalizes excessive stroke modifications. Strokes with low $Q(s_i)$ values are either adjusted or discarded.

\textbf{Adaptive Stroke Merging:} Similarly, we can have a step for handling conflicting strokes. For example, conflicting strokes are where neural predictions deviate significantly from heuristic rules. We can merge the two using a weighted blending function:
\begin{equation}
s_i^{\text{final}} = \omega s_i^* + (1 - \omega) s_j^*,
\end{equation}
where $\omega$ is computed based on local structural similarity between neighboring strokes $s_i^*$ and $s_j^*$. This approach is particularly effective in preserving smooth transitions in NPR techniques like hatching and painterly rendering \cite{hertzmann2003survey}.

Finally, this step produces a finalized, balanced stroke sequence $\mathcal{S}^*$, ensuring artistic adaptability while maintaining stroke coherence. The refined strokes are then rendered and post-processed in the final stage.

\subsection{Step 5: Rendering and Post-Processing}

This is the final step in the algorithm \ref{alg:hybrid_stroke_general}. After stroke sequences have been optimized, the final rendering phase generates the output image while ensuring perceptual coherence. We can employ variations of the post-processing depending on the requirements, for example to further refine the result by removing artifacts and improving visual smoothness, particularly in video-based NPR applications.

\textbf{Sequential Stroke Rendering:} To maintain artistic coherence, strokes are rendered sequentially based on their priority and structure. Given the final stroke sequence $\mathcal{S}^*$, strokes are drawn in a controlled order:
\begin{equation}
I^* = \sum_{i=1}^{N} R(s_i^*),
\end{equation}
where $R(s_i^*)$ represents the rendering function applied to each stroke. Techniques such as opacity-based layering and texture blending can be used to enhance realism.


\textbf{Post-Processing:} To refine the stylized output, additional post-processing operations can be applied, including:
\begin{itemize}
    \item \textbf{Edge enhancement:} Reinforces contours for clearer structure.
    \item \textbf{Noise reduction:} Uses smoothing filters to remove undesired artifacts.
    \item \textbf{Color harmonization:} Adjusts stroke blending to improve consistency.
\end{itemize}
The final output is a fully rendered stroke-based image, having both artistic expressiveness and structural coherence. The post-processed result is suitable for applications in digital painting, stylized animations, and interactive NPR rendering.

This modular framework serves as a foundation for future NPR implementations, allowing researchers to explore different deep neural models, reinforcement learning policies, and hybrid optimization strategies.


\section{Conclusion \& Future Work}
Current neural models struggle with precise control over perceptual attributes like abstraction, style, and stroke-level manipulation. We argue that such perceptual attributes can be best addressed through hybrid approaches. Hybrid approaches enable users to dynamically adjust stylization parameters (Algorithm \ref{alg:hybrid_stroke_general}). As such, suitable versions of Algorithm \ref{alg:hybrid_stroke_general} can be integrated into interactive tools to allow for stroke refinement, dynamic parameter adjustment, and seamless blending of neural content with manual intervention. In light of this, manual intervention highlights the need for improved NPR evaluation metrics. Specifically, NPR evaluation remains challenging due to inherent subjectivity. Thus, there is a need for the development of computational metrics, informed by perceptual studies and artist feedback.

Neural stylization methods often introduce artifacts or fail to generalize to unfamiliar styles. Conversely, style reconstruction techniques, while effective, are computationally demanding. Therefore, future research should prioritize developing efficient neural architectures and lightweight models. Furthermore, the limited availability of curated, artist-created datasets hinders the generalization capabilities of neural models. To address this, expanding datasets with greater diversity and employing domain-specific pretraining could significantly enhance performance. Additionally, some models may replicate biases present in training data. Future work should integrate computational aesthetics, interdisciplinary insights from art, psychology, and ethics, and human-in-the-loop evaluation frameworks. Resolution of the aforementioned challenges is paramount to the advancement of NPR systems.

\bibliographystyle{plain} 
\bibliography{references}  

\begin{thebibliography}{1}

\bibitem{dev2013mobile}
Kapil Dev.
\newblock Mobile expressive renderings: The state of the art.
\newblock {\em IEEE Computer Graphics and Applications}, 33(3):22--31, 2013.

\bibitem{hertzmann2003survey}
A~Hertzmann.
\newblock A survey of stroke-based rendering.
\newblock {\em IEEE Computer Graphics and Applications}, 23(4):70--81, 2003.

\bibitem{hertzmann2010non}
Aaron Hertzmann.
\newblock Non-photorealistic rendering and the science of art.
\newblock In {\em Proceedings of the 8th International Symposium on Non-Photorealistic Animation and Rendering}, pages 147--157, 2010.

\bibitem{jing2019neural}
Yongcheng Jing, Yezhou Yang, Zunlei Feng, Jingwen Ye, Yizhou Yu, and Mingli Song.
\newblock Neural style transfer: A review.
\newblock {\em IEEE transactions on visualization and computer graphics}, 26(11):3365--3385, 2019.

\bibitem{kyprianidis2012state}
Jan~Eric Kyprianidis, John Collomosse, Tinghuai Wang, and Tobias Isenberg.
\newblock State of the ''art'': A taxonomy of artistic stylization techniques for images and video.
\newblock {\em IEEE transactions on visualization and computer graphics}, 19(5):866--885, 2012.

\bibitem{lansdown1995expressive}
John Lansdown and Simon Schofield.
\newblock Expressive rendering: A review of nonphotorealistic techniques.
\newblock {\em IEEE Computer Graphics and Applications}, 15(3):29--37, 1995.

\bibitem{liu2021paint}
Songhua Liu, Tianwei Lin, Dongliang He, Fu~Li, Ruifeng Deng, Xin Li, Errui Ding, and Hao Wang.
\newblock Paint transformer: Feed forward neural painting with stroke prediction.
\newblock In {\em Proceedings of the IEEE/CVF international conference on computer vision}, pages 6598--6607, 2021.

\bibitem{liu2023painterly}
Xiao-Chang Liu, Yu-Chen Wu, and Peter Hall.
\newblock Painterly style transfer with learned brush strokes.
\newblock {\em IEEE Transactions on Visualization and Computer Graphics}, 30(9):6309--6320, 2023.

\bibitem{ryan1956speed}
Thomas~A Ryan and Carol~B Schwartz.
\newblock Speed of perception as a function of mode of representation.
\newblock {\em The American journal of psychology}, 69(1):60--69, 1956.

\end{thebibliography}

\begin{IEEEbiography}{Kapil Dev}{\,}is a senior lecturer at RMIT University, Vietnam. His research interests include computer graphics, human-computer interaction, and computer vision. Dev has an PhD in Computer Science from Lancaster University. Contact him at: kapil.saini@hotmail.com.

\end{IEEEbiography}

\end{document}